\documentclass[a4paper,fleqn,usenatbib]{mnras}

\usepackage{newtxtext,newtxmath}

\usepackage[T1]{fontenc}
\usepackage{ae,aecompl}

\usepackage{graphicx}	
\usepackage{amsmath}	
\usepackage{amssymb}	

\newcommand{\kms}{km s$^{-1}$}

\title[Environmental Effects on Galaxy Spin]{
A Study of Environmental Effects on Galaxy Spin Using MaNGA Data}

\author[J. C. Lee et al.]{
Jong Chul Lee,$^{1}$\thanks{E-mail: jclee@kasi.re.kr}
Ho Seong Hwang$^{2}$ 
and Haeun Chung$^{3,4}$
\\
$^{1}$Korea Astronomy and Space Science Institute, 
776 Daedeokdae-ro, Yuseong-gu, Daejeon 34055, Korea\\
$^{2}$Quantum Universe Center, Korea Institute for Advanced Study,
85 Hoegiro, Dongdaemun-gu, Seoul 02455, Korea\\
$^{3}$School of Physics, Korea Institute for Advanced Study,
85 Hoegiro, Dongdaemun-gu, Seoul 02455, Korea\\
$^{4}$Astronomy Program, Department of Physics and Astronomy,
Seoul National University, Gwanak-gu, Seoul 08826, Korea
}

\date{Accepted XXX. Received YYY; in original form ZZZ}

\pubyear{2017}

\begin{document}
\label{firstpage}
\pagerange{\pageref{firstpage}--\pageref{lastpage}}
\maketitle

\begin{abstract}
We investigate environmental effects on galaxy spin 
  using the recent public data of MaNGA integral field spectroscopic survey 
  containing $\sim$2800 galaxies. 
We measure the spin parameter of 1830 galaxies 
  through the analysis of two-dimensional stellar kinematic maps 
  within the effective radii, and obtain their large- 
  (background mass density from 20 nearby galaxies) 
  and small-scale (distance to and morphology of the nearest neighbour galaxy) 
  environmental parameters for 1529 and 1767 galaxies, respectively.
We first examine the mass dependence of galaxy spin,
  and find that the spin parameter of early-type galaxies decreases
  with stellar mass at log (M$_*/$M$_{\sun}$) $\gtrsim$ 10,
  consistent with the results from previous studies.
We then divide the galaxies into three subsamples
  using their stellar masses
  to minimize the mass effects on galaxy spin.
The spin parameters of galaxies in each subsample
  do not change with background mass density,
  but do change with distance to and morphology
  of the nearest neighbour.
In particular, the spin parameter of late-type galaxies decreases 
  as early-type neighbours approach within the virial radius.
These results suggest that the large-scale environments
  hardly affect the galaxy spin,
  but the small-scale environments
  such as hydrodynamic galaxy-galaxy interactions
  can play a substantial role in determining galaxy spin.
\end{abstract}




\section{Introduction} \label{intro}

Measurement of galaxy spin is important to understand 
  the physical conditions when the galaxies are formed. 
Since \citet{pee69} first proposed the tidal torque theory 
  to explain the origin of the angular momentum of galaxies, 
  the galaxy spin has been extensively
  studied by numerical simulations
  \citep[e.g.][]{too77,her92,cox06,naa14,cho17}.
In the context of hierarchical galaxy formation paradigm,
  it is naturally expected that
  early-type galaxies are dispersion dominated,
  while late-type galaxies are rotation dominated.
However, this simple morphology-spin relation has been recently revised
  by integral field spectroscopic (IFS) observations,
  which reveal that a substantial fraction of early-type galaxies
  are significantly rotating \citep[e.g.][]{cap07,ems07,kra08}.

The early-type galaxies can be divided into two groups, 
  fast and slow rotators, based on the spin parameter that indicates 
  the relative importance of stellar rotation and dispersion 
  and that is a proxy of specific angular momentum \citep[see][]{ems07}.
The fast and slow rotators are also called regular and non-regular rotators,
  respectively; the regular rotators are galaxies with kinematic discs,
  while the non-regular rotators are galaxies with
  little rotation, irregular rotation, kinematically decoupled cores 
  or counter-rotating discs \citep[see][]{kra11}.
There are more slow rotators in more massive galaxies
  \citep[e.g.][]{bro17,oli17,van17a,vea17}.
This result is consistent with the observation that 
  the spin parameter of galaxies usually decreases with stellar mass;
  this could be understood from an analytical relation 
  among spin parameter, angular momentum and total mass
  \citep[e.g.][]{fal80,rom12,cor16},
  and is now reproducible in hydrodynamic simulations \citep{lag17,cho18}.

The well-known morphology-density relation
  (i.e. a higher fraction of early-type galaxies in denser regions) 
  has been also revisited in terms of galaxy spin 
  \citep[e.g.][]{mac07,cer08,cap11,fog14}.
For example, the slow rotator fraction seems to increase 
  with local density, but such an environmental dependence 
  disappears when galaxy mass is fixed (e.g. \citealt{bro17,vea17,gre18a}; but see also \citealt{cho18}). 
Thanks to recent systematic IFS surveys including 
  SAMI (Sydney-AAO Multi-object IFS, \citealt{bry15}) and
  MaNGA (Mapping Nearby Galaxies at APO, \citealt{bun15}),
  we are now able to study the details of environmental dependence of galaxy spin 
  with larger samples.

Here, we use MaNGA, one of the largest IFS datasets currently available, 
  to study the environmental effects on galaxy spin. 
This paper is different from previous studies 
  based on similar MaNGA datasets in several ways \citep{gre18a,gre18b}. 
Greene et al. focused on $\sim$500 early-type central and 
  satellite galaxies in galaxy groups, 
  and found no environmental dependence of spin parameter and 
  no difference in spin parameter between central and satellite galaxies at fixed mass. 
In this study, we increase the sample size 
  using all the galaxies in the latest data release of the MaNGA survey 
  (there are 984 early- and 846 late-type galaxies in the final sample of this study) 
  and estimate diverse environmental parameters 
  representing large- and small-scale environments. 
We then examine the environmental dependence of spin parameters 
  for several galaxy samples divided by morphological type and stellar mass.
The structure of this paper is as follows.
We describe the data and sample in Section \ref{data}.
The methods to determine the galaxy spin and 
  the environmental parameters are explained in Section \ref{anal}.
Our results are presented and discussed in Section \ref{result}.
We summarize the results in Section \ref{summ}.
Throughout this paper, we adopt a flat $\Lambda$CDM cosmology with
  $H_0$ = 100 $h$ km s$^{-1}$ Mpc$^{-1}$,
  $\Omega_{\Lambda}$ = 0.7 and $\Omega_m$ = 0.3.
In stellar masses and absolute magnitudes, 
  the terms of ``$h^{-2}$'' and ``$-$5 log $h$'' are omitted, respectively.

\section{Data and Sample} \label{data}

MaNGA is an IFS survey project,
  one of the fourth-generation Sloan Digital Sky Survey 
  (SDSS-IV) programs \citep{bla17},
  to obtain two-dimensional spectroscopic data
  of $\sim$10,000 nearby galaxies by 2020 
  (see \citealt{bun15} and \citealt{yan16} for an overview).
The MaNGA spectrograph \citep{dro15} uses different-sized hexagonal bundles
  of 2 arcsec diameter fibres
  (from 12$\arcsec$ diameter for a 19-fibre bundle
  to 32$\arcsec$ for a 127-fibre bundle)
  and cover the spectral range of 3600--10300 \AA\
  with a resolution R $\sim$ 2000 
  (FWHM $\sim$ 2.6 \AA; $\sigma$ $\sim$ 70 \kms).
Each target is observed with three dithers
  for continuous spatial coverage without leaving unobserved gaps.
The synthetic maps provided by the data reduction pipeline \citep{law16}
  have the pixel size 0$\arcsec$.5 and effective spatial resolution
  (FWHM$_{\rm seeing}$) $\sim2\arcsec$.5 in $r$-band.

The physical parameters of MaNGA galaxies are adopted
  from many catalogues. 
For example, the structure parameters and stellar mass estimates 
  are taken from the NASA Sloan Atlas catalogue\footnote{http://nsatlas.org}.
The structural parameters including
  effective radius R$_{\rm e}$, minor-to-major axis ratio b/a
  and position angle are measured with elliptical Petrosian apertures.
The stellar mass M$_*$ is derived from
  the spectral energy distribution fit
  to the SDSS $ugriz$-band photometry
  with the \citet{bru03} stellar population model
  and \citet{cha03} initial mass function.
The right ascension, declination, redshift, and $r$-band magnitude
  of supplementary galaxies used
  for determining environmental parameters of MaNGA galaxies
  (see Section \ref{envir}) are drawn
  from the SDSS Data Release 12 catalogue \citep{ala15}.
The absolute magnitude M$_r$ is calculated with
  the Galactic reddening correction \citep{sch98},
  $K$-correction \citep[shifted to $z$ = 0.1]{bla07}
  and evolution correction \citep{teg04}.

The galaxy morphology information (early/late) is adopted from
  the Korea Institute for Advanced Study (KIAS) DR7 value-added 
  galaxy catalogue\footnote{http://astro.kias.re.kr/vagc/dr7/} \citep{cho10}, 
  the Galaxy Zoo catalogues \citep{lin11,wil13} and 
  our additional visual inspection. 
The KIAS DR7 value-added galaxy catalogue contains 
  the morphological information of $\sim$708,000 galaxies. 
The galaxies are divided into early (ellipticals and lenticulars) and 
  late (spirals and irregulars) types based on their locations 
  in the ($u-r$) colour versus ($g-i$) colour gradient space and 
  in the $i$-band concentration index space \citep{par05}. 
The resulting morphological classification has completeness and 
  reliability reaching 90 per cent. 
The thirteen astronomers in the KIAS group performed 
  an additional visual check of the SDSS $gri$ colour images of 
  the galaxies misclassified by the automated scheme. 
In this procedure, they revised the types of blended or merging galaxies, 
  blue but elliptical-shaped galaxies and dusty edge-on spirals. 
The Galaxy Zoo is a project to provide morphological information of
  SDSS galaxies from hundreds of thousands of volunteers \citep{lin08}. 
The citizen scientists determine galaxy morphology 
  using the $gri$ colour images and the resulting catalogue 
  provides a probability of a galaxy to be early- or late-type. 
We use Galaxy Zoo 1 and 2 catalogues that include $\sim$900,000 
  and $\sim$304,000 galaxies, respectively. 
The agreement of galaxy morphology between the KIAS VAGC and 
  the Galaxy Zoo catalogues is $\sim$81 per cent for the galaxies 
  in the SDSS main galaxy sample at m$_{r} <$ 17.77 mag. 
We take the morphology of KIAS VAGC when the two classifications differ. 
A detailed understanding of the cause of the difference 
  between the two classifications is beyond the scope of the paper. 
We also performed visual classification for the 2367 galaxies 
  at m$_{r} <$ 17.77 mag that are not included in 
  KIAS DR7 VAGC and in Galaxy Zoo catalogues. 
All the three authors visually inspected the SDSS $gri$ colour images 
  and determined the morphology. 
To reduce possible bias introduced by different morphological classifications, 
  we also visually inspected all the MaNGA galaxies and 
  their nearest neighbours (see Section \ref{envir}) in this study. 
In the result, among 636,447 galaxies with morphological information
  at m$_{r} <$ 17.77 mag, 92, 7 and 1 per cent of the classifications 
  are taken from KIAS VAGC, Galaxy Zoo and visual inspection, respectively.

\begin{figure}
\includegraphics[width=\columnwidth]{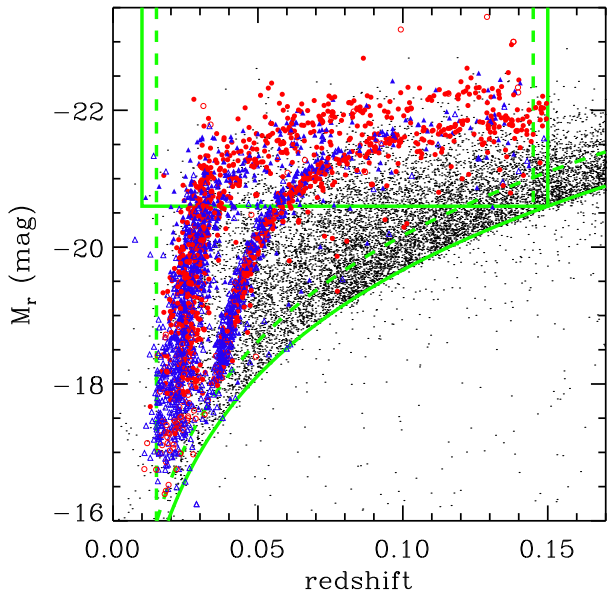}
\caption{The $r$-band absolute magnitude versus redshift diagram.
Red circles and blue triangles indicate early- and late-type
MaNGA galaxies, respectively.
Filled (open) symbols are the MaNGA galaxies
with (without) measured spin parameters.
Black dots denote the spectroscopic sample of galaxies in SDSS DR12.
Only one per cent of galaxies are shown for better visibility.
Straight lines define the volume-limited sample
to estimate large-scale background densities.
Bottom curve means the apparent magnitude limit (m$_r =$ 17.77 mag)
of the SDSS main galaxy sample. 
When investigating the environmental dependence of spin parameters,
we use target galaxies within the region surrounded by dashed lines
to reduce the volume incompleteness effects (see Section \ref{envir}).
\label{fig-sample}}
\end{figure}

The IFS data we use are from the second public release of the MaNGA
  as part of SDSS DR14 \citep{dr14},
  which include 2778 galaxies.
If there are repeated observations for the same target,
  we choose the one with a higher median S/N (signal-to-noise ratio)
  of spaxels within R$_{\rm e}$.
Among the 2726 unique galaxies, we could measure spin parameters
  for 1830 galaxies (see Section \ref{spin}).
Figure \ref{fig-sample} shows the redshift and M$_r$ distributions
  of these MaNGA galaxies.
There are two sequences corresponding to
  the primary and secondary targets that are designed
  to cover up to 1.5 and 2.5 R$_{\rm e}$, respectively.

\section{Analysis} \label{anal}

\subsection{Spin parameter} \label{spin}

\begin{figure*}
\includegraphics[width=\linewidth]{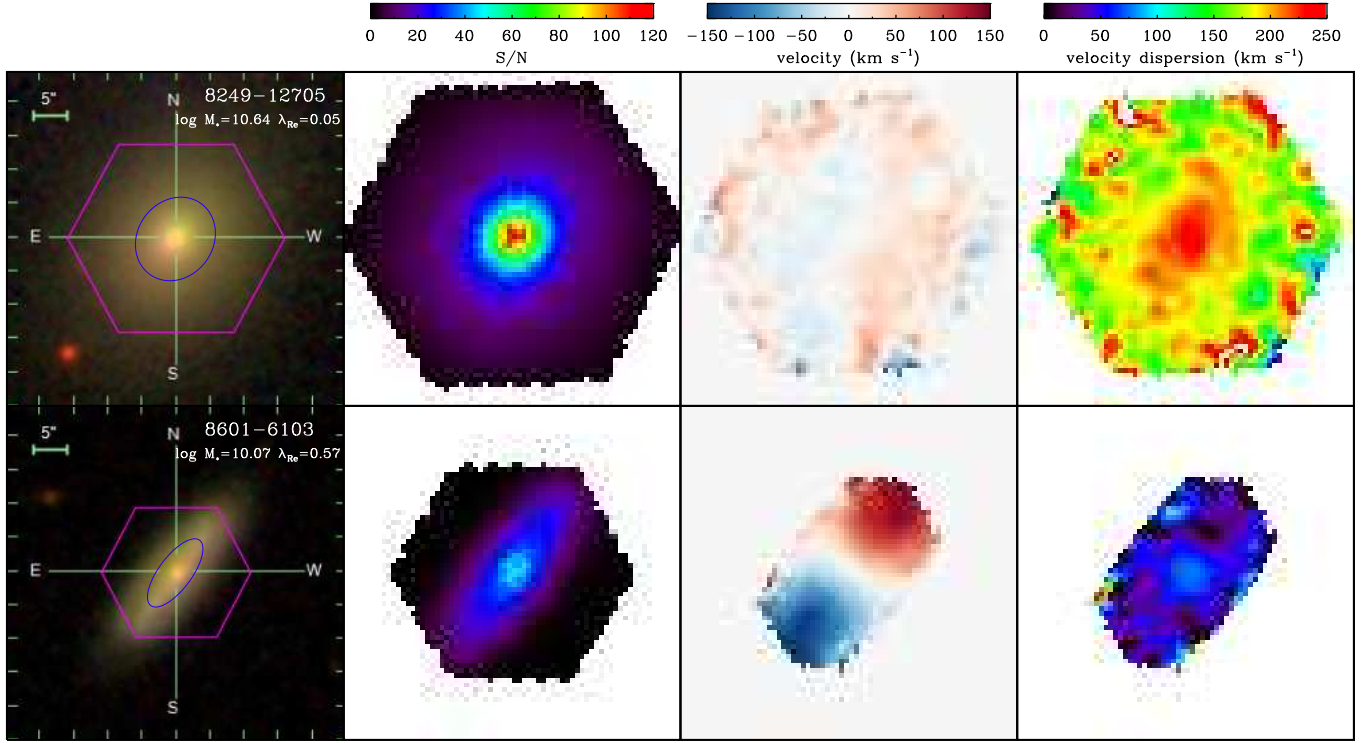}
\caption{
Example of kinematic maps for early-type (top row) 
and late-type (bottom row) galaxies.
The SDSS $gri$-band composite images are shown in the 1st column
with their MaNGA ID, stellar mass and spin parameter.
The pink hexagon and blue ellipse superimposed on the image denote
the spectrograph's field-of-view and 
aperture to measure spin parameter.
The S/N, velocity and velocity dispersion maps are listed 
in the 2nd, 3rd and 4th columns, respectively.
The size of SDSS image is 50$\arcsec \times$ 50$\arcsec$ and 
that of others is 35$\arcsec \times$ 35$\arcsec$. \label{fig-exam}}
\end{figure*}

\begin{figure}
\includegraphics[width=\columnwidth]{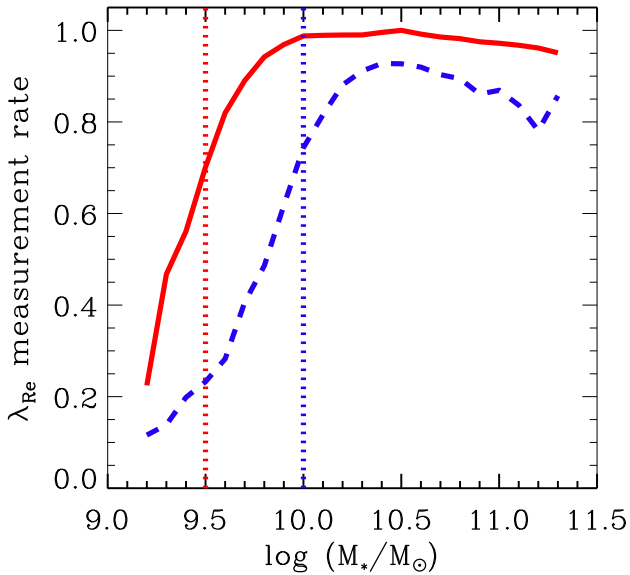}
\caption{Spin parameter measurement rate as a function of stellar mass
for early- (red solid line) and late-type (blue dashed line) galaxies.
The vertical dotted lines denote the mass limits 
of 70 per cent completeness for each type. \label{fig-comp}}
\end{figure}

To extract stellar kinematic information
  (line-of-sight velocity and its dispersion)
  from the MaNGA data cubes, 
  we use the penalized pixel fitting code (pPXF) of \citet{cap04}
  and \citet{vaz10} synthetic stellar population model  
  (156 templates covering 3540.5--7409.6 \AA, 1--18 Gyr 
  and [M/H] = from $-$1.71 to $+$0.22 
  with a resolution of FHWM = 2.51 \AA).
We run the code on the spectra of spaxels with median S/N $>$ 2 and
  with the number fraction of flagged elements
  from the MaNGA pipeline $<$ 20 per cent\footnote{
  Among the 2726 unique galaxies, there are 331 galaxies that have spaxels 
  with $\geq$ 20 per cent of flagged elements within the effective radii. 
  The numbers of galaxies with $>$1, $>$10 and $>$30 per cent of spaxels 
  masked due to the criterion of flagged elements are 247, 34 and one, respectively. 
  Therefore, there is only one galaxy whose spin parameter could not be measured 
  mainly because of this. 
  On average, the MaNGA galaxies have two (0.6 per cent) spaxels 
  with $\geq$ 20 per cent of flagged elements within the effective radii.}
  after de-redshifting and masking major emission lines.
The options used for the run are velocity dispersion shape of ``Gaussian''
  and additive Legendre polynomial order of ``10''.
As a sanity check,
  we construct a subsample of 100 galaxies that are randomly selected 
  by considering morphological type, stellar mass and spectrum S/N. 
We then run the code on their spectra 
  with MILES stellar library of \citet{san06} (985 templates).
The resulting maps are consistent with 
  those from the Vazdekis model templates 
  for the spaxels with (intrinsic) velocity dispersion $\geq$ 40 \kms
  (see Appendix \ref{appenA}),
  but the fitting speed is much slower \citep[see also][]{van17b}.
The velocity dispersion smaller than 40 \kms\ is very uncertain 
  because it is seriously affected by 
  the instrumental resolution limit \citep{pen16}.

Following \citet{ems07,ems11},
  we measure the luminosity-weighted, 
  dimensionless spin parameter $\lambda_{\rm R}$ 
  by integrating the kinematic quantities within the ellipse:
\begin{equation}
\lambda_{\rm R} = \frac{<R|V|>}{<R\sqrt{V^2+\sigma^2}>} = 
\frac{\sum_{i=1}^{N}F_iR_i|V_i|}{\sum_{i=1}^{N}F_iR_i\sqrt{V_i^2+\sigma_i^2}},
\end{equation}
  where $N$ is the total number of spaxels within the radius of R.
  $F_{i}$, $R_{i}$, $V_{i}$, and $\sigma_{i}$ are the $r$-band flux,
  semi-major axis, stellar velocity, and velocity dispersion
  of the $i$th spaxel, respectively.
The $\lambda_{\rm R}$ is calculated at R = 0.1--3.0 R$_{\rm e}$
  in steps of 0.1 R$_{\rm e}$.
We use only the spaxels with median S/N $\geq$ 10 and
  with velocity dispersion $\geq$ 40 km/s, 
  not affected by adjacent objects\footnote{
  The adjacent objects are those in the SDSS photometric catalogue 
  within the IFS field of view except the target galaxy. 
  We mask the spaxels within the radii containing 90 per cent of 
  Petrosian flux in the $r$-band (denoted as ``petroR90$_{r}$'' 
  in the SDSS catalogue) of these objects. 
  Among the 2726 unique galaxies, 
  we could not measure spin parameters for 29 galaxies ($\sim$1.1 per cent) 
  because they have close companions; we had to mask many spaxels within their effective radii.
  To properly obtain two-dimensional maps of velocity and velocity dispersion 
  for these galaxies, 
  a detailed modelling with multiple kinematic components should be performed in the pPXF procedure. 
  We therefore simply exclude these galaxies from the analysis in this study.}
When the number fraction of spaxels
  used for the $\lambda_{\rm R}$ measurement is larger than 70 per cent 
  and the radius for the $\lambda_{\rm R}$ measurement
  is larger than FWHM$_{\rm seeing}$ (to mitigate beam-smearing effects),
  we consider the $\lambda_{\rm R}$ estimate reliable.
Among several estimates of $\lambda_{\rm R}$
  at different radii for a given galaxy,
  we choose the one at the radius close to R$_{\rm e}$
  as the $\lambda_{\rm Re}$.
To minimize the effects of different aperture size 
  on the spin parameter measurement \citep[see also][]{van17a}, 
  we focus only on 1830 galaxies whose $\lambda_{\rm Re}$ is determined 
  in a narrow range R = 0.8--1.2 R$_{\rm e}$ 
  from the sample of 2726 unique MaNGA galaxies. 
The spin parameter for 89 per cent of the final sample 
  is determined exactly at R = R$_{\rm e}$.
The comparison of measured spin parameters for MaNGA galaxies 
  from two different methods suggests that the uncertainty 
  in the spin parameter measurement is about 0.09 \citep{gre18b}.
Figure \ref{fig-exam} shows two representative cases
  demonstrating the process of measuring spin parameter.
Most galaxies excluded from the final sample of 1830 galaxies 
  are less massive late-type galaxies
  that have many spaxels with velocity dispersion $<$ 40 \kms
  in the outer regions.
In Figure \ref{fig-comp}, we plot the completeness of spin parameter measurements 
  as a function of morphology and mass 
  (i.e. fraction of galaxies with measured spin parameters). 
The completeness drops quickly for less massive late-type galaxies
  that have many spaxels with velocity dispersion $<$ 40 \kms
  in the outer regions.
It also slightly decreases in the high mass end 
  mainly because the spatial coverage of IFS data is not large enough 
  to reach the effective radii of some galaxies. 
However, the completeness of spin parameters remains above 70 per cent for 
  early- and late-type galaxies at log (M$_*$/M$_\odot) >$ 9.5 and 10.0, respectively.

\subsection{Galaxy environment} \label{envir}

We consider two kinds of environmental parameters:
  a background mass density $\rho_{20}$ as a large-scale environment
  and a projected distance to the nearest neighbour galaxy R$_{\rm n}$
  (and its morphological type) as a small-scale environment.
Full details of how to calculate these parameters are
  provided in \citet{par08} and \citet{par09}.

Briefly, $\rho_{20}$ is a large-scale mass density measured with
  20 nearby galaxies weighted by their mass and distance
  over a few Mpc scale.
We calculate $\rho_{20}$ for each MaNGA galaxy
  using its 20 closest galaxies (in three-dimensional space)
  in the volume-limited sample shown in Figure \ref{fig-sample}
  (0.01 $\leq z \leq$ 0.15 and M$_r \geq -20.60$ mag)
  following the formula of \citet{par08}.
We use $\rho_{20}$ divided by
  the mean mass density of the universe, adopted from \citet{par08}:
  $<$$\rho$$>$ = 0.0223 $\pm$ 0.0005 ($\gamma L$)$_{-20}$ 
  where ($\gamma L$)$_{-20}$ is the mass of a late-type galaxy 
  with M$_{r}$ = $-$20 mag.

We also identify the nearest neighbour of a target galaxy
  that is the closest on the projected sky among galaxies
  with the absolute magnitude brighter than M$_{r,\rm target}$ + 0.5 mag
  and the radial velocity difference less than
  600 \kms\ for early-type targets or
  400 \kms\ for late-type targets.
The velocity difference limits are determined from the rms velocity difference 
  between a target galaxy and its neighbours, 
  which is nearly constant out to the projected separation of 50 kpc. 
Those values can cover most close neighbors as seen in figure 1 of \citet{par08}.
The R$_{\rm n}$ is normalized by
  the nearest neighbour's virial radius r$_{\rm vir,n}$.
The virial radius of a galaxy is defined by the projected radius 
  where the mean mass density is 200 times 
  the critical density of the universe ($\rho_{c}$) as
  r$_{\rm vir}$ = ($3\gamma L/4\pi)^{1/3}(200\rho_{c})^{-1/3}$
  where $L$ is the galaxy luminosity and $\gamma$ is the mass-to-light ratio. 
We assume that the mass-to-light ratio of early-type galaxies is twice 
  as large as that of late-type galaxies at the same absolute magnitude M$_r$ 
  (see section 2.3 of \citealt{par09} for more details).
The virial radii of galaxies with M$_r$ = $-$19.5, $-$20.0 and $-$20.5 mag
  are 260, 300 and 350 $h^{-1}$ kpc for early types,
  and 210, 240 and 280 $h^{-1}$ kpc for late types, respectively.

If a target galaxy is close to the SDSS survey boundary,
  its environmental parameters may be incorrectly determined.
To avoid this problem, when investigating the environmental effects
  in Section \ref{main},
  we use only target galaxies at z = 0.015--0.145\footnote{
  The redshift margin of 0.005 (i.e. 1500 \kms\ buffer) is large enough 
  to avoid the volume incompleteness effect 
  because our velocity difference limits to identify nearest neighbours 
  are smaller than 1500 \kms. 
  This margin is also acceptable for $\rho_{20}$ 
  when we consider the distribution of velocity differences 
  between a target and its 20 closest galaxies. 
  Among the 1529 galaxies with  used in this study, 
  only 4.1 per cent (0.7 per cent) of galaxies have the 20th (10th) neighbour 
  with the velocity difference $>$ 1500 \kms. 
  The maximum velocity difference to the 20th (10th) galaxy is 2170 (1830) \kms. 
  If we adopt a larger redshift margin, 
  the number of galaxies for the analysis becomes smaller.} 
  and with the projected distance to the sky boundary
  larger than the half of the distance
  to the most distant galaxy used for each parameter
  (20-th galaxies for $\rho_{20}$ and
  nearest neighbour for R$_{\rm n}$).
In addition, to ensure the spectroscopic completeness
  of neighbouring galaxies
  when we identify the nearest neighbour
  among those with M$_{r,\rm neighbour} \leq$ M$_{r,\rm target}$ + 0.5 mag,
  we restrict our analysis to the MaNGA galaxies 
  with m$_r \leq$ 17.77 $-$ 0.5 mag
  where 17.77 mag is the apparent magnitude limit of the SDSS main galaxy sample.
We remove, respectively, 16 and three per cent of MaNGA galaxies
  with unreliable $\rho_{20}$ and R$_{\rm n}$ estimates from our analysis.
In the result, there are 1529 and 1767 galaxies 
  with reliable $\rho_{20}$ and R$_{\rm n}$ estimates, respectively.

To examine the effects of different choices in calculating 
  environmental parameters on the results, 
  we conduct the following experiments. 
We estimate the background densities using 10 and 30 nearby galaxies 
  instead of 20, and still find no dependence of 
  spin parameter on these parameters (see Section \ref{main}).
When we use the velocity criterion of $\Delta$v $<$ 500 \kms\ 
  for both early- and late-type galaxies to identify their nearest neighbours, 
  the spin parameter of late-type galaxies at log (M$_*$/M$_\odot) =$ 10.0--10.5 
  still decreases as early-type neighbours approach. 
The significance level of this change is similar to 
  the one with the original criterion (i.e. $\Delta$v $<$ 600 and 400 \kms\ 
  for early- and late-type galaxies, respectively).
These suggest that our conclusions do not change with different conditions 
  in calculating environmental parameters.
On the other hand, if we select the nearest neighbour among the galaxies 
  brighter than M$_{r,\rm target}$ + 1 mag instead of M$_{r,\rm target}$ + 0.5 mag, 
  the effect of neighbour galaxy on spin parameter becomes insignificant.
This is expected because the inclusion of less (gravitationally/hydrodynamically)
  effective neighbors weakens the signal.
It is also noted that the reliability of the smoothed density field 
  based on 20 neighbouring galaxies used in this study (i.e. $\rho_{20}$) 
  is demonstrated in appendix of \citet{son16}. 
In addition, extensive comparisons among different environmental parameters 
  including $\rho_{20}$ and other conventional measures are provided in \citet{mul12}.

The spectroscopic completeness of the SDSS data is poor 
  for bright galaxies at m$_{r} <$ 14.5 mag 
  because of the problems of saturation and cross-talk in the spectrograph, 
  and for the galaxies located in high-density regions 
  including galaxy clusters because of fibre collisions. 
We therefore supplement the SDSS data with redshifts from the literature 
  for the galaxies with m$_{r} <$ 17.77 mag (see \citealt{hwa10} for more detail). 
The resulting completeness at m$_{r} <$ 17.77 mag 
  for the SDSS DR12 data in this study becomes, on average, 
  $\sim$95 per cent and its spatial variation is smaller than 5 per cent 
  except for certain small regions 
  (see figure 1 of \citealt{par09} and figure 3 of \citealt{cho10}). 
The overall high completeness and its small spatial variation ensure that 
  the bias in the measurements of background density 
  introduced by spectroscopic incompleteness is very small. 
The spectroscopic incompleteness also can affect 
  the identification of genuine neighbours; 
  the nearest neighbour can be seriously misidentified 
  if the completeness is very low. 
The fibre collision issue can make this problem worse. 
Our previous Monte Carlo experiment shows that 
  the fraction of misidentified nearest neighbours can be 
  $\sim$50 and $\sim$5 per cent when the sample completeness is 50 and 95 per cent, 
  respectively \citep{hwa09}. 
The overall completeness in our sample is $\sim$95 per cent, 
  and we confirm that the completeness does not change with the projected distance to the target galaxy. 
In conclusion, the effect of SDSS spectroscopic incompleteness on 
  the environment measurements is not significant in this study 
  thanks to the high completeness of our SDSS sample,
  supplemented by the redshifts from the literature.

\section{Results and Discussion} \label{result}

\subsection{Mass dependence of spin parameter} \label{mass}

\begin{figure}
\includegraphics[width=\columnwidth]{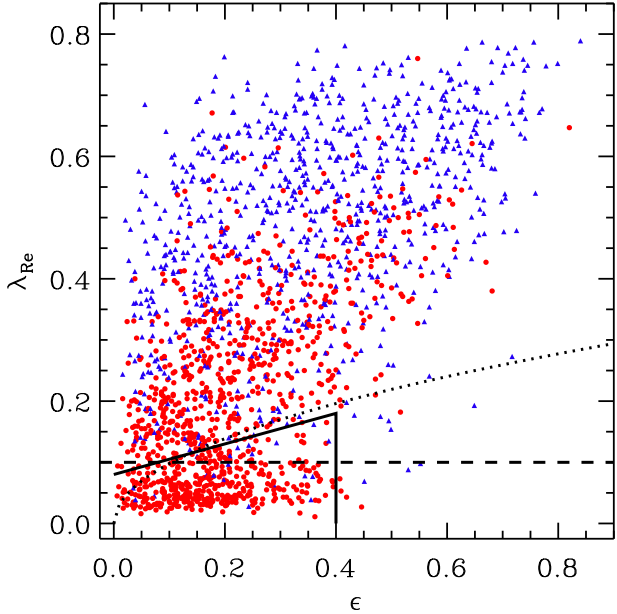}
\caption{Spin parameter versus ellipticity diagram.
Symbols are the same as Figure \ref{fig-sample}.
Different lines indicate the criteria to distinguish between
slow and fast rotators,
suggested by \citet[dashed]{ems07}, \citet[dotted]{ems11}
and \citet[solid]{cap16}. \label{fig-class}}
\end{figure}

\begin{figure}
\center
\includegraphics[width=\columnwidth]{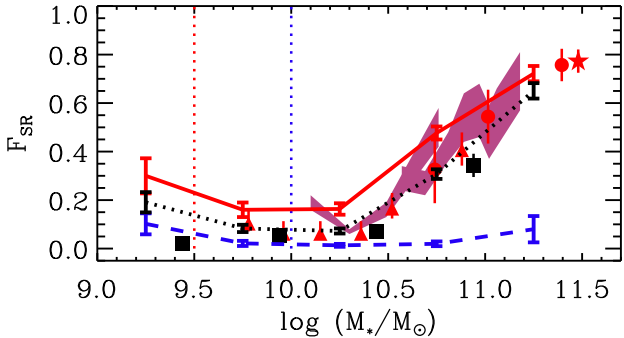}
\caption{
Number fraction of slow rotators, selected by the \citet{cap16} criteria,
in 0.5 dex mass bins. Red solid, blue dashed and black dotted lines are
for early, late and both types, respectively.
The vertical dotted lines indicate the mass limits of 70 per cent completeness.
Error bars are derived from the binomial statistics.
The other symbols indicate the slow rotator fractions in previous studies.
Red triangles \citep{bro17}, red circles \citep{oli17}, red star \citep{vea17}
and pink shaded regions \citep{gre18b} are for the slow rotator fraction 
of early types.
Black squares \citep{van17b} are for that of early plus late types.
All stellar masses are based on $h$ = 1 and Chabrier initial mass function.
\label{fig-frac}}
\end{figure}

\begin{figure*}
\center
\includegraphics[width=0.8\linewidth]{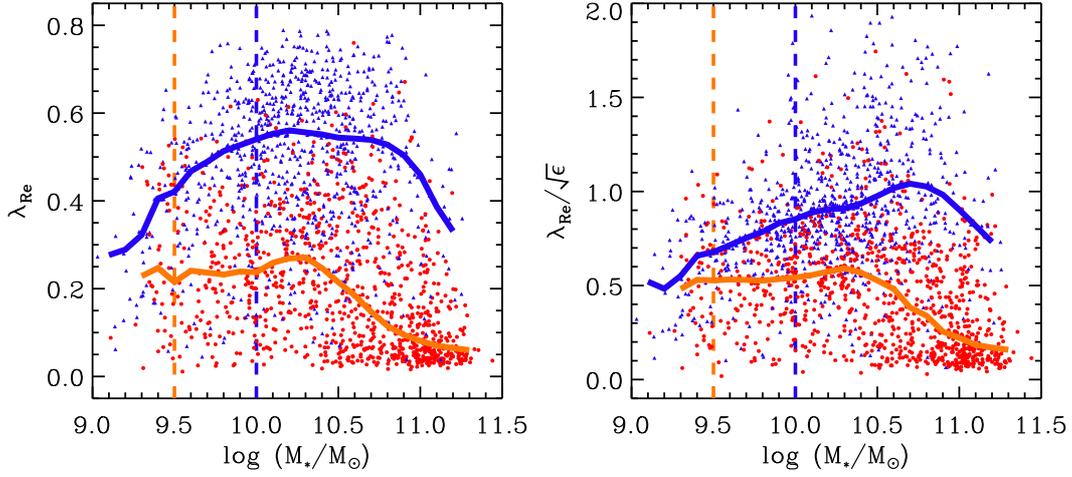}
\caption{Spin parameter $\lambda_{\rm Re}$ (left) and
$\lambda_{\rm Re}$/$\sqrt{\epsilon}$ (right)
as a function of stellar mass.
Orange and blue solid lines represent
sliding medians for early and late types, respectively.
The vertical dashed lines indicate the 70 per cent completeness limits.
\label{fig-mass}}
\end{figure*}

Figure \ref{fig-class} shows the spin parameter $\lambda_{\rm Re}$
  against ellipticity $\epsilon$ ($=1-$b/a) of MaNGA galaxies.
There are several methods to distinguish slow rotators
  from fast rotators in this domain:
  $\lambda_{\rm Re} < 0.1$ \citep[dashed]{ems07},
  $\lambda_{\rm Re} < 0.31 \sqrt{\epsilon}$ \citep[dotted]{ems11}
  and $\lambda_{\rm Re} < 0.25 \epsilon + 0.08$ \& $\epsilon < 0.4$
  \citep[solid]{cap16}.
If we adopt the criterion of \citet{cap16},
  the fractions of slow rotators in our sample (F$_{\rm SR}$)
  are 0.394 $\pm$ 0.016 and 0.025 $\pm$ 0.005
  for early and late types, respectively.

Figure \ref{fig-frac} shows the stellar mass dependence of 
  slow rotator fraction F$_{\rm SR}$ in our sample 
  together with the results in the literature. 
The F$_{\rm SR}$ increases with mass dramatically in early types 
  at log (M$_*$/M$_\odot) \gtrsim$ 10,
  consistent with other results \citep[e.g.][]{bro17,oli17,van17a,vea17,gre18b}.
It is difficult to examine the variation of
  F$_{\rm SR}$ for late types
  because of small numbers of slow rotators in each mass bin.
On the other hand, the slow rotator fractions in this study 
  appear slightly higher than those in other studies.
This difference could be because low spatial resolution and sampling 
  in MaNGA survey make spin parameters be measured small. 
Actually, when we use the galaxies at z = 0.02--0.04 
  with a relatively high (physical) spatial resolution, 
  we obtain a significantly smaller fraction of slow rotators 
  (e.g. F$_{\rm SR}$ for early-type galaxies at log (M$_*/$M$_{\sun}$) = 10.5--11.0 
  becomes 0.343 $\pm$ 0.057 from 0.477 $\pm$ 0.027). 
This fraction is more similar to the other results than 
  the one using all the galaxies. 
We discuss this issue in more detail 
  in Appendix \ref{appenB} (see also \citealt{gre18b}).
 
The left panel of Figure \ref{fig-mass} demonstrates that
  the spins of both early- and late-type galaxies decrease
  with stellar mass at the massive end.
Interestingly, $\lambda_{\rm Re}$ of early and late types
  appears roughly constant at log (M$_*$/M$_\odot) \lesssim$ 10.2 and 10.8,
  respectively.
The increase of spin parameter with stellar mass for late types 
  at log (M$_*$/M$_\odot) \lesssim$ 10 may not be reliable 
  because a substantial fraction of less massive late types
  especially with high $\lambda_{\rm Re}$ are excluded in this study,
  as mentioned in Section \ref{spin}.
To better understand the different mass dependence of $\lambda_{\rm Re}$
  on galaxy morphology,
  it is necessary to obtain more measurements of $\lambda_{\rm Re}$ of
  less massive late types using deeper observations
  with higher spectral resolutions.

It should be noted that $\lambda_{\rm Re}$ is a projected quantity,
  thus $\lambda_{\rm Re}$ could be underestimated in face-on galaxies
  \citep[e.g.][]{cap07,ems11}.
To take into account the inclination effect
  on the measured $\lambda_{\rm Re}$,
  we divide $\lambda_{\rm Re}$
  by $\sqrt{\epsilon}$ \citep[e.g.][]{ems11,cor16}
  in the right panel of Figure \ref{fig-mass}.
The mass dependence of $\lambda_{\rm Re}$/$\sqrt{\epsilon}$
  is similar to that of $\lambda_{\rm Re}$ for early types.
However, late types show a different pattern;
  $\lambda_{\rm Re}$/$\sqrt{\epsilon}$ does not have such a plateau
  and does increase with stellar mass at log (M$_*$/M$_\odot) \lesssim$ 10.7,
  different from the case of $\lambda_{\rm Re}$
  in the left panel.
Because galaxies are expected to have random angles of inclination
  regardless of their stellar masses,
  the inclination correction should not change
  the mass dependence of spin parameter
  before and after the correction.
This indicates that the simple correction method
  using the ellipticity may produce a bias for late-type galaxies;
  for example, the ellipticity in this study
  is from the ratio of semi-major/minor axes,
  which can be significantly affected by
  the presence of bulge
  that is nothing to do with inclination
  (see also \citealt{hol12,wei14}).
We therefore use the original $\lambda_{\rm Re}$
  without the inclination correction for further analysis.

\subsection{Environmental dependence of spin parameter} \label{main}

\begin{figure*}
\center
\includegraphics[width=\linewidth]{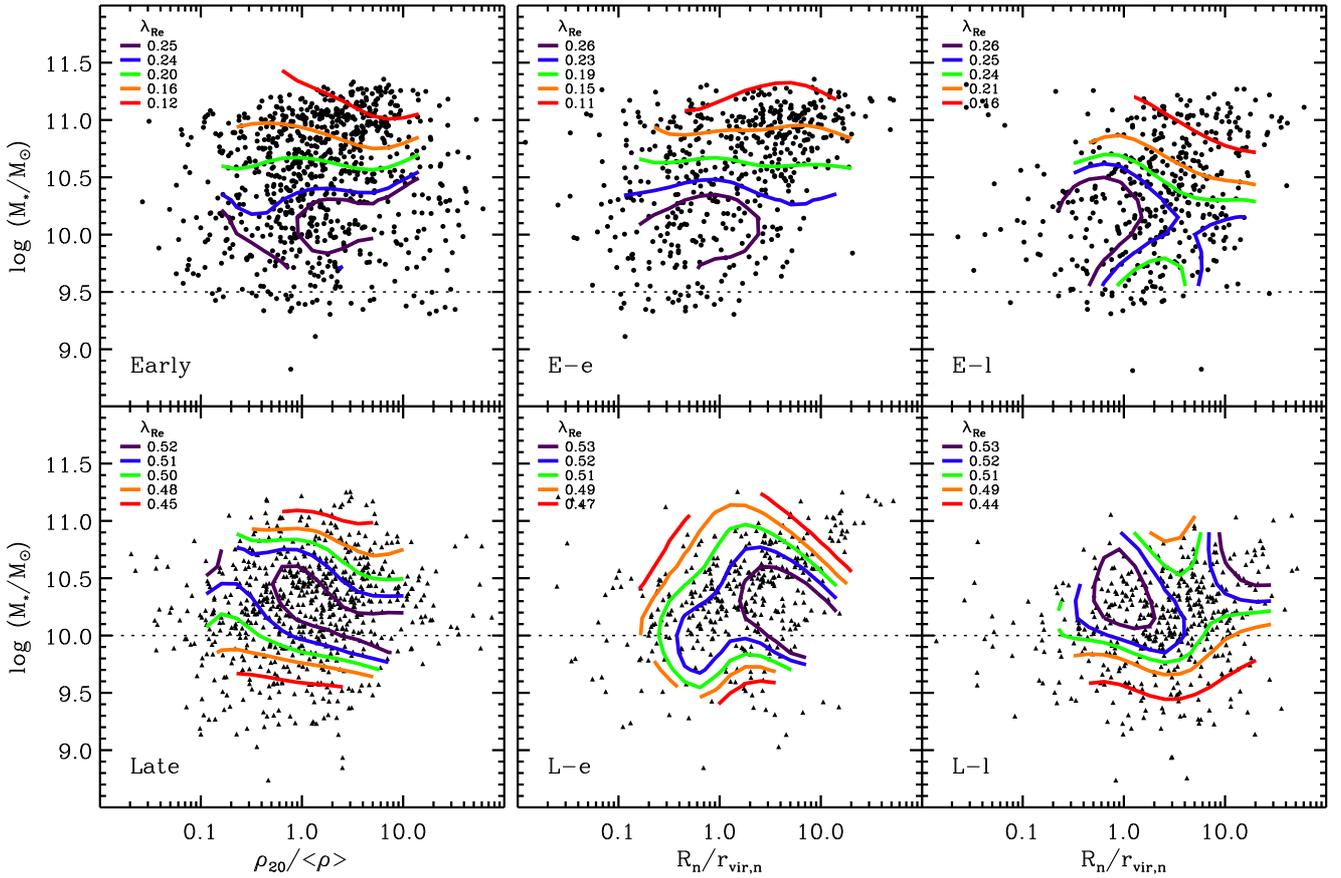}
\caption{Spin parameter contours in the stellar mass
and environmental parameter space.
The left panels are for the large-scale background density.
The middle and right panels are for the (projected) pair separation
between target galaxies and their nearest neighbours.
Four cases are shown:
early-type target galaxies with an early-type neighbour (top middle, E-e),
late types with an early-type neighbour (bottom middle, L-e),
early types with a late-type neighbour (top right, E-l), and
late types with a late-type neighbour (bottom right, L-l). 
The horizontal dotted lines indicate the mass limits 
of 70 per cent completeness. \label{fig-contour}}
\end{figure*}

\begin{figure*}
\center
\includegraphics[width=\linewidth]{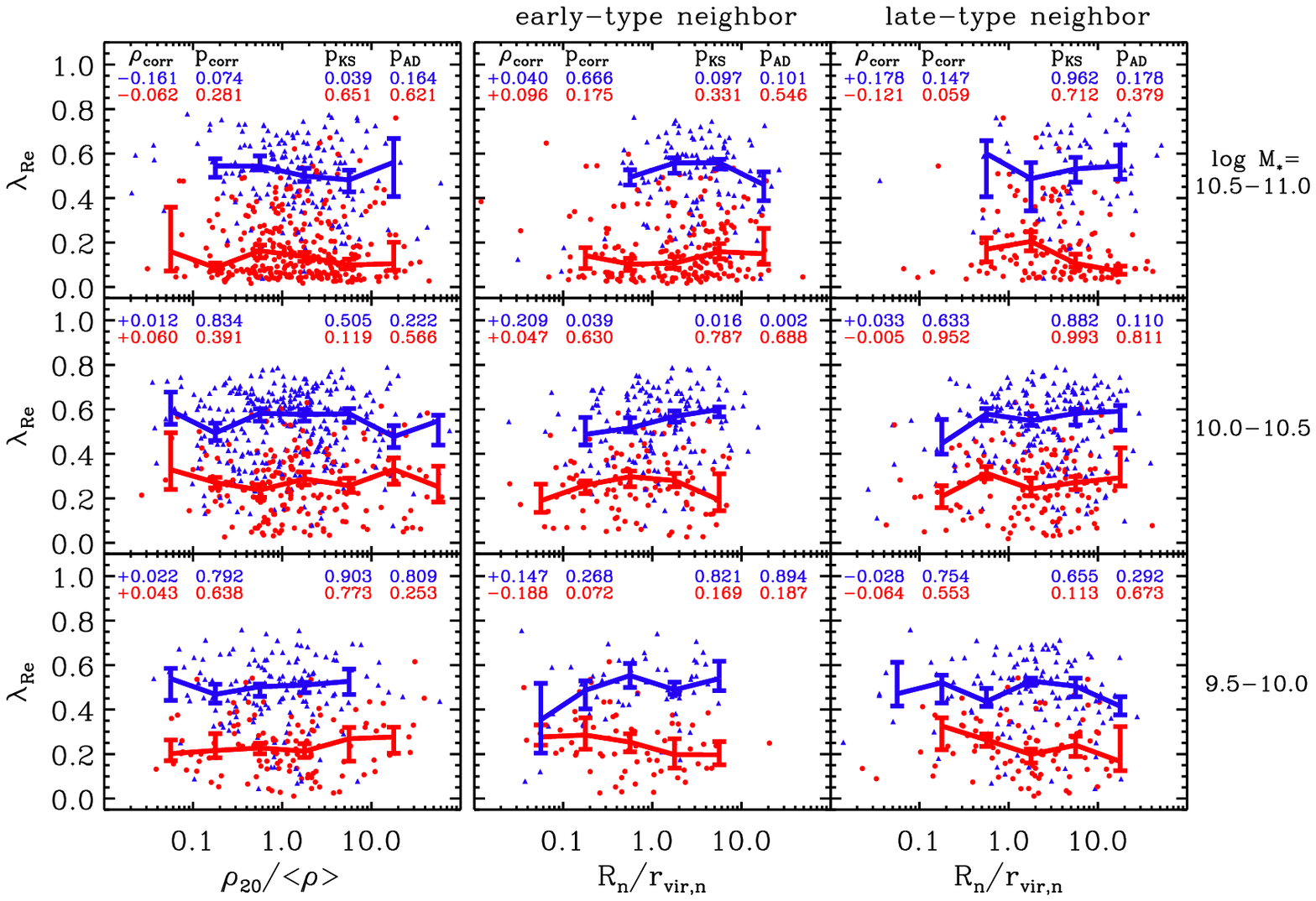}
\caption{Dependence of spin parameter on the environmental parameters.
The environmental parameters are the same as Figure \ref{fig-contour}.
The target galaxies in the top, middle and bottom panels are limited to
their stellar mass ranges of log (M$_{*}$/M$_{\sun}$) = 10.5--11.0,
10.0--10.5 and 9.5--10.0, respectively.
The numbers in the top left corner of each panel denote
the Spearman's rank correlation coefficient and
the probability of obtaining the correlation by chance.
The numbers in the top right corner mean
the $p$-values from the K-S and A-D $k$-sample tests between galaxies
with the environmental parameter $\leq$ 1 and $>$ 1.
The first and second rows are for late and early types.
The median values of spin parameter in 0.5 dex bins of
environmental parameters are also presented with sampling errors;
red and blue lines are for early and late types, respectively.
\label{fig-relation}}
\end{figure*}

Figure \ref{fig-contour} displays the $\lambda_{\rm Re}$
  dependence on stellar mass and environmental parameters.
The contours in the left panels are mostly horizontal. 
This means that the spin parameter of early and late types 
  strongly depends on stellar mass, 
  but does hardly depend on background density. 
On the other hand, the middle and right panels show that
  the contours are not always horizontal (except in the top middle panel). 
This suggests that the spin parameters are affected by 
  both stellar mass and nearest neighbour. 
We also note that the contour patterns differ
  between middle and right panels.
This indiactes that the spin parameter changes differently 
  depending on the morphological type of neighbour.

To better examine the environmental dependence of $\lambda_{\rm Re}$
  after minimizing the mass effect,
  we present $\lambda_{\rm Re}$ as a function of
  environmental parameters
  for three mass bins of
  log (M$_{*}$/M$_{\sun}$) = 10.5--11.0 (high),
  10.0--10.5 (intermediate) and 9.5--10.0 (low)
  in Figure \ref{fig-relation}.
We use several statistical tools to check how meaningful the correlation
  between $\lambda_{\rm Re}$ and environmental parameter is:
  the Spearman correlation test, the Kolmogorov-Smirnov (K-S) test
  and the Anderson-Darling (A-D) $k$-sample test.
The probability of obtaining the given correlation by chance
  in the Spearman test\footnote{If the value is smaller than 0.05,
  in general, the correlation is regarded as significant
  above 2$\sigma$ confidence level.} indicates that
  the dependence of $\lambda_{\rm Re}$ on $\rho_{20}$/$<$$\rho$$>$
  is not significant regardless of morphological type and stellar mass
  (see numbers in the left corner of each panel).
However, the $\lambda_{\rm Re}$ variation 
  with R$_{\rm n}$/r$_{\rm vir,n}$
  is found with a 2.1 $\sigma$ level
  when late-type targets have early-type neighbours
  in the intermediate mass bin (i.e. blue curve in the centre panel).
There is also a hint (1.9 $\sigma$) of $\lambda_{\rm Re}$ variation
  with R$_{\rm n}$/r$_{\rm vir,n}$
  when early-type targets have late-type neighbours
  in the high mass bin (i.e. red curve in the top right panel).
The decrease of $\lambda_{\rm Re}$ of late-type galaxies
  when early-type neighbour is within its viral radius
  is confirmed (2.4--3.1 $\sigma$) in the intermediate mass bin
  with the K-S and A-D $k$-sample tests
  by comparing the galaxies at R$_{\rm n} \leq$ r$_{\rm vir,n}$
  and at R$_{\rm n} >$ r$_{\rm vir,n}$
  (see numbers in the right corner of each panel).

A decrease of $\lambda_{\rm Re}$ of late-type galaxies
  as early-type neighbours approach and 
  a possible increase of $\lambda_{\rm Re}$ of early-type galaxies
  as late-type neighbours approach in this study
  imply that galaxy spin can be influenced by 
  not only gravitational but also hydrodynamic
  interactions with neighbouring galaxies.
This is consistent with the results
  from the {\small EAGLE} hydrodynamic cosmological simulation
  of \citet{lag18} that the specific angular momentum of
  galaxies is decreased ($\sim$30 per cent) and increased ($\sim$10 per cent)
  via gas-poor and gas-rich mergers, respectively
  (but see also \citealt{pen17}).
In fact, galaxy mergers are widely accepted to be responsible for
  the formation of early-type slow rotators \citep[e.g.][]{duc11,sme18}
  through dynamical friction
  that efficiently moves material with high angular momentum
  to the outer regions \citep[see][]{zav08} and/or 
  through cold disc gas removal by ram pressure and tidal stripping
  \citep[see][]{may06}.
On the other hand, \citet{lag18} explain the increase of
  angular momentum by the centrally concentrated, young stars
  with high rotational speed, which are connected to 
  interaction-induced gas inflow and accretion.
The star formation activity enhanced by interactions
  with late-type neighbours has indeed been observed
  in several studies
  \citep[e.g.][]{par09,hwa10,hwa11,dav15,cao16}.
These suggest that the combination of stellar and gas kinematics
  as well as the consideration of different stellar populations
  is important for better understanding
  the hydrodynamic effects on galaxy spin \citep[see also][]{cor14}.

Among the 1830 MaNGA galaxies analysed in this study,
  there are 138 galaxies in Abell 2199 supercluster region \citep{hwa12}.
The superclusters of galaxies are excellent laboratories
  for studying the environmental dependence of galaxy properties
  thanks to its wide coverage of galaxy environment \citep[e.g.][]{lee15}.
We thus separately examine the dependence of galaxy spin of A2199 galaxies
  on projected clustercentric distance, which could be supplementary to
  the large-scale background density we use, $\rho_{20}$/$<$$\rho$$>$.
However, we do not find that the galaxy spin changes with
  clustercentric distance,
  compatible with the results of \citet{bro17} and \citet{gre18a}.
To draw a strong conclusion,
  it will be helpful to construct a large sample of cluster galaxies
  with IFS data \citep[see][]{owe17} and
  to use more efficient tools like a phase-space diagram of
  clustercentric distance versus line-of-sight velocity \citep[e.g.][]{oma13,rhe17}.

\section{Summary} \label{summ}

We measure the spin parameter of 1830 MaNGA galaxies 
  from the analysis of two-dimensional stellar spectra of galaxies
  to study their environmental dependence. 
We use the background mass density $\rho_{20}$ and
  the distance to the nearest neighbour galaxy R$_{\rm n}$
  as large- and small-scale environmental parameters, respectively.
Our main results are as follows.

\begin{enumerate}
\item Among 2726 MaNGA galaxies (1109 early and 1617 late types),
      we measure the spin parameters of 984 early- and 
      846 late-type galaxies.
      We could not measure the spin parameter for the remaining galaxies
      because of our strict selection criteria for reliable measurements and 
      of the low spectral resolution of MaNGA data.
\item The spin parameter of early-type galaxies decreases 
      with stellar mass at log (M$_*/$M$_{\sun}$) $\gtrsim$ 10,
      similar to the results from previous studies.
\item The spin parameter of both early- and late-type galaxies 
      shows no dependence on the background mass density 
      when stellar mass is fixed.
      This suggests that the large-scale environment does not 
      affect the galaxy spin.
\item The spin parameter of late-type galaxies appears to decrease 
      as early-type neighbours approach within the virial radius.
      This indicates a non-negligible impact of small-scale environment
      on galaxy spin,
      possibly via hydrodynamic interactions with neighbouring galaxies.
\end{enumerate}

\section*{Acknowledgements}

We thank the anonymous referee for helpful comments 
that improved the manuscript.
We thank Kwang-Il Seon and the KIAS Center for Advanced Computation
for providing computing resources.
J.C.L. is a member of Dedicated Researchers for Extragalactic AstronoMy
(DREAM) in Korea Astronomy and Space Science Institute (KASI).

Funding for the Sloan Digital Sky Survey IV has been provided by
the Alfred P. Sloan Foundation, the U.S. Department of Energy Office of
Science, and the Participating Institutions. SDSS-IV acknowledges
support and resources from the Center for High-Performance Computing at
the University of Utah. The SDSS web site is www.sdss.org.

SDSS-IV is managed by the Astrophysical Research Consortium for the 
Participating Institutions of the SDSS Collaboration including the 
Brazilian Participation Group, the Carnegie Institution for Science, 
Carnegie Mellon University, the Chilean Participation Group, 
the French Participation Group, Harvard-Smithsonian Center for Astrophysics, 
Instituto de Astrof\'isica de Canarias, The Johns Hopkins University, 
Kavli Institute for the Physics and Mathematics of the Universe 
(IPMU) / University of Tokyo, Lawrence Berkeley National Laboratory, 
Leibniz Institut f\"ur Astrophysik Potsdam (AIP),  
Max-Planck-Institut f\"ur Astronomie (MPIA Heidelberg), 
Max-Planck-Institut f\"ur Astrophysik (MPA Garching), 
Max-Planck-Institut f\"ur Extraterrestrische Physik (MPE), 
National Astronomical Observatories of China, New Mexico State University, 
New York University, University of Notre Dame, 
Observat\'ario Nacional / MCTI, The Ohio State University, 
Pennsylvania State University, Shanghai Astronomical Observatory, 
United Kingdom Participation Group,
Universidad Nacional Aut\'onoma de M\'exico, University of Arizona, 
University of Colorado Boulder, University of Oxford, 
University of Portsmouth, University of Utah, University of Virginia, 
University of Washington, University of Wisconsin, Vanderbilt University, 
and Yale University.







\appendix

\section{Comparison between Vazdekis and MILES models} \label{appenA}

\begin{figure*}
\includegraphics[width=140mm]{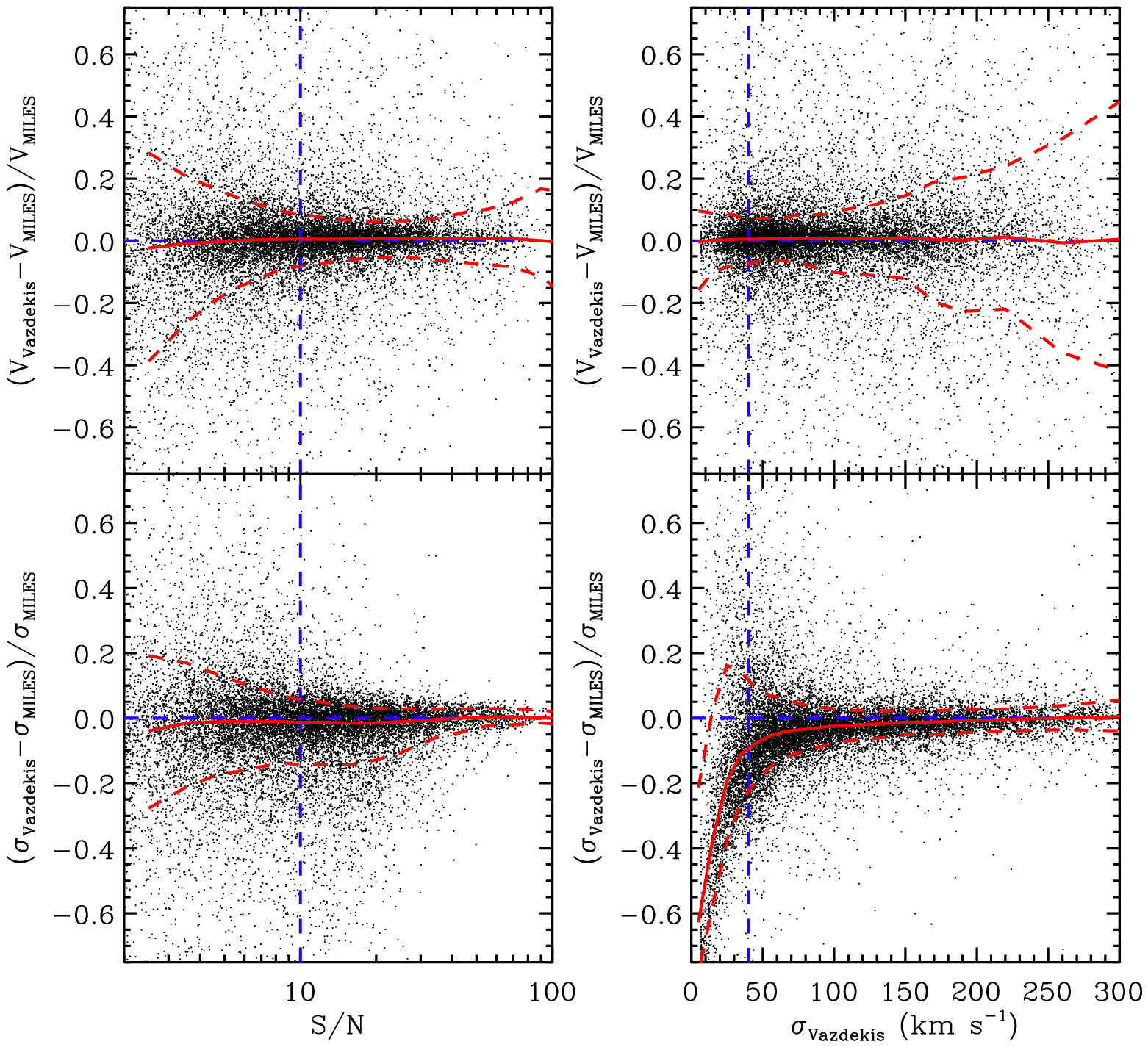}
\caption{
Comparison of pPXF outputs between Vazdekis and MILES models
as a function of S/N (left) and of velocity dispersion from the Vazdekis model 
(right).
For better visibility, only ten per cent of spaxels are presented with 
sliding median (red solid) and 68 per cent enclosure (red dashed) lines.
Blue vertical lines indicate the lower limits of our selection criteria. \label{fig-appenA1}}
\end{figure*}

We use 100 galaxies in the test sample of Section \ref{spin}
  to examine the differences of the pPXF outputs
  between the Vazdekis and MILES models.
Figure \ref{fig-appenA1} shows that the velocity and velocity dispersion from the Vazdekis model 
  generally agree with those from the MILES library in the ranges of our criteria for spaxels to analyse
  (i.e. S/N $\geq$ 10 and $\sigma_{\rm Vazdekis}$ $\geq$ 40 \kms).

\section{Spatial resolution effect on spin parameters} \label{appenB}

\begin{figure*}
\includegraphics[width=140mm]{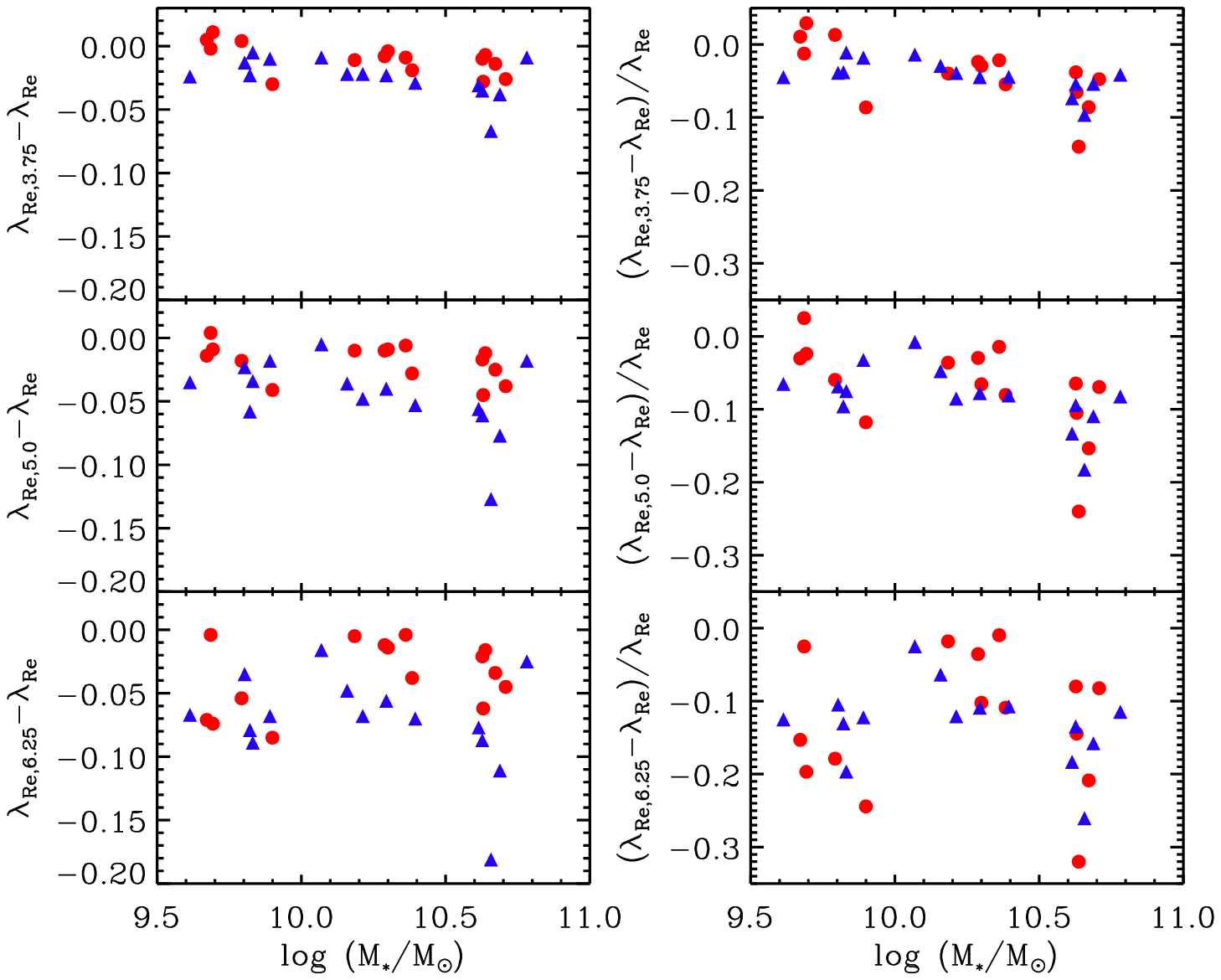}
\caption{
The difference in spin parameter measurement between smoothed 
and unsmoothed IFS data (left: $\Delta\lambda_{\rm Re}$,
right: $\Delta\lambda_{\rm Re}$/$\lambda_{\rm Re}$).
The ``$\lambda_{\rm Re,number}$'' means the spin parameter derived 
from the Gaussian-convolved IFS data with effective FWHM = number (arcsec).
Red circles and blue triangles are for early and late types, respectively.
\label{fig-appenB1}}
\end{figure*}

\begin{figure*}
\includegraphics[width=140mm]{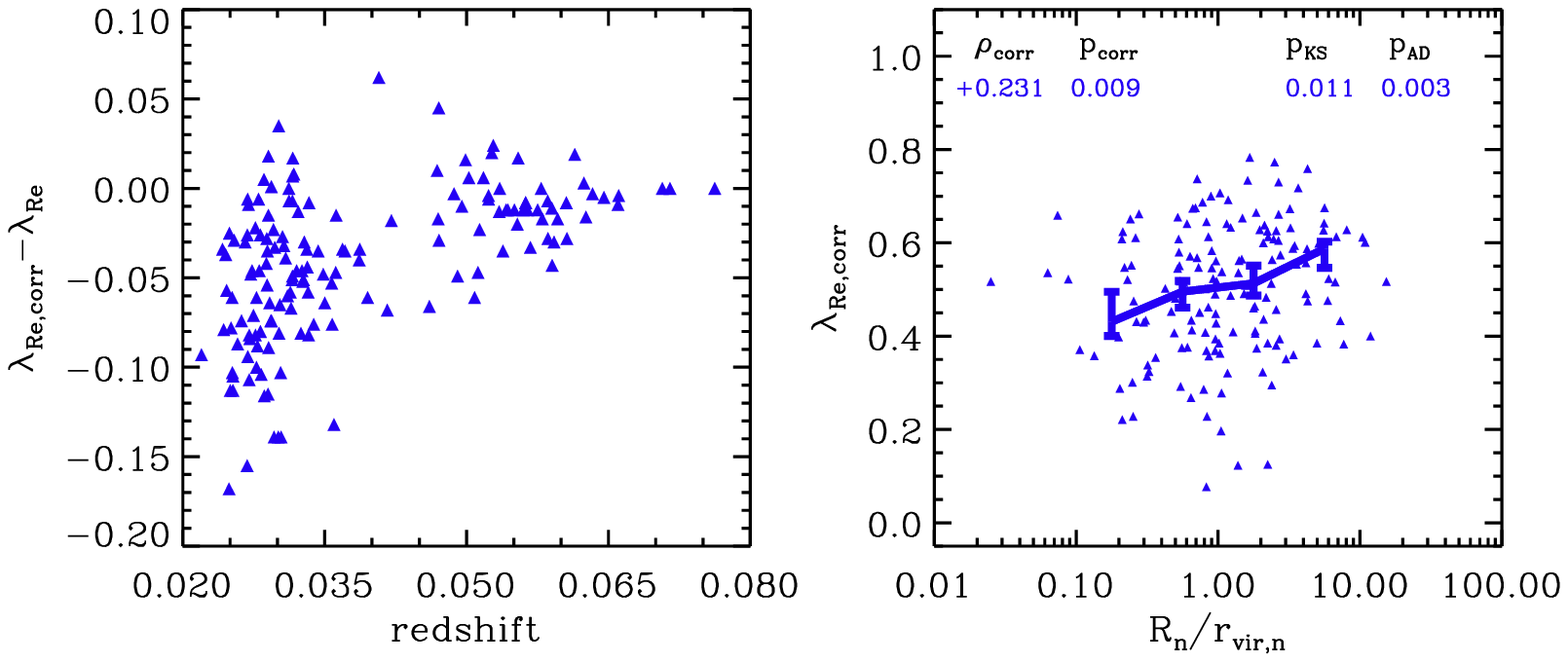}
\caption{
(Left) The difference between resolution-corrected ($\lambda_{\rm Re,corr}$)
and uncorrected ($\lambda_{\rm Re}$) spin parameters 
as a function of redshift
for 156 late-type galaxies in the intermediate mass bin with 
an early-type neighbour.
(Right) The resolution-corrected spin parameter 
as a function of distance to the nearest neighbour.
The numbers are the results from the statistical tests 
as in Figure \ref{fig-relation}. \label{fig-appenB2}}
\end{figure*}

To investigate the spatial resolution effects on the measurement of spin parameter, 
  we first select 30 nearby galaxies at z = 0.02--0.04 
  by considering morphological type and stellar mass 
  (i.e. 5 early types and 5 late types at each mass bin). 
We then generate the Gaussian-convolved IFS data with effective 
  FWHM = 3$\arcsec$.75, 5$\arcsec$.0 and 6$\arcsec$.25 
  using the original IFS data with typical FWHM of 2$\arcsec$.5, 
  and measure the spin parameters. 
The resulting spin parameters are compared with the original values 
  in Figure \ref{fig-appenB1}.
In most cases, the spin parameters are measured lower for lower spatial resolution data. 
The left panels show that the median differences in spin parameter estimates 
  between the original data and the Gaussian-convolved ones for early types are 
  0.009, 0.014 and 0.034 for the convolved IFS data with 
  FWHM = 3$\arcsec$.75, 5$\arcsec$.0 and 6$\arcsec$.25, respectively. 
For late types, the median differences are 0.030, 0.040 and 0.068.
The $\Delta\lambda_{\rm Re}$ differs between early and late types, 
  but the difference becomes much smaller 
  when comparing fractional changes in the right panels 
  (0.038, 0.065 and 0.109 for early types vs. 0.041, 0.081 and 0.122 for late types).
The top and middle panels show that the spin parameter variation appears 
  larger in more massive galaxies, 
but this mass dependency is not obvious in the bottom panels.

Because the resolution effect on spin parameter estimate 
  is not negligible as shown in Figure \ref{fig-appenB1}, 
  we further examine whether or not our main conclusion is affected by this issue. 
We focus on the 161 late-type galaxies with early-type neighbours 
  in the intermediate mass bin (log (M$_*/$M$_{\sun}$) = 10.0--10.5) 
  where the spin parameter change is most significant. 
The physical spatial resolution for these galaxies ranges from 1.2 to 3.5 kpc. 
We then match the spatial resolution by constructing 
  Gaussian-convolved IFS data with effective FWHM = 3.5 kpc 
  and measure the spin parameters for 156 galaxies.
We exclude five galaxies from this measurement because their (Gaussian-convolved) 
  effective radii exceed the IFS field of view. 
We plot the change in spin parameter estimate due to the resolution correction 
  as a function of redshift in the left panel of Figure \ref{fig-appenB2}. 
The right panel shows that our conclusion on the neighbor dependence of 
  spin parameter remains unchanged even if we use the resolution-corrected spin parameters; 
  the spin parameter of late-type galaxies decreases as early-type neighbours approach, 
  and its statistical significance is still larger than 2.5 $\sigma$.
It should be noted that the amount of change in spin parameter estimate 
  due to the resolution correction does not depend on the distance to nearest neighbor. 
Nevertheless, we do not think that it is practical to apply such a resolution correction to the entire sample of galaxies 
  especially for massive galaxies 
  because the range for the physical resolution is too broad.
This means that many galaxies will be removed during this correction procedure. 
Therefore, we use the spin parameters measured without resolution correction.

\bsp	
\label{lastpage}
\end{document}